\documentclass[twocolumn,showpacs,amsmath,amssymb,aps,prb]{revtex4}

\usepackage{graphicx}% Include figure files
\usepackage{multirow}
\usepackage{color}
\usepackage{dcolumn}% Align table columns on decimal point
\usepackage{bm}% bold math
\usepackage{subfigure}
\newcommand{\bra}[1] {\langle #1 |}
\newcommand{\ket}[1] {| #1 \rangle}

\newcommand{\cg}[6] {\left(\begin{array}{cc|c} #1 & #3 & #5  \\ #2 & #4 & #6 \end{array}\right)}
\newcommand{\rdm}[3]{\bra{#1}| #2 | \ket{#3}}
\newcommand{\me}[3]{\bra{#1} #2 \ket{#3}}

\begin{document}

\title{Theory of the ground state spin of the NV$^-$ center in diamond: I. Fine structure, hyperfine structure, and interactions with electric, magnetic and strain fields}

\author{M.W. Doherty$^1$, F. Dolde$^2$, H. Fedder$^2$, F. Jelezko$^{2,3}$, J. Wrachtrup$^2$, N.B. Manson$^4$, and L.C.L. Hollenberg$^1$}

\affiliation{$^1$ School of Physics, University of Melbourne, Victoria 3010, Australia \\
$^2$ 3$^{\mathrm{rd}}$ Institute of Physics and Research Center SCOPE, University Stuttgart, Pfaffenwaldring 57, D-70550 Stuttgart, Germany \\
$^3$ Institut f$\mathrm{\ddot{u}}$r Quantenoptik, Universit$\mathrm{\ddot{a}}$t Ulm, Ulm D-89073, Germany \\
$^4$ Laser Physics Centre, Research School of Physics and Engineering, Australian National University, Australian Capital Territory 0200, Australia}

\date{20 July 2011}

\begin{abstract}
The ground state spin of the negatively charged nitrogen-vacancy center in diamond has been the platform for the recent rapid expansion of new frontiers in quantum metrology and solid state quantum information processing. In ambient conditions, the spin has been demonstrated to be a high precision magnetic and electric field sensor as well as a solid state qubit capable of coupling with nearby nuclear and electronic spins. However, in spite of its many outstanding demonstrations, the theory of the spin has not yet been fully developed and there does not currently exist thorough explanations for many of its properties, such as the anisotropy of the electron g-factor and the existence of Stark effects and strain splittings. In this work, the theory of the ground state spin is fully developed for the first time using the molecular orbital theory of the center in order to provide detailed explanations for the spin's fine and hyperfine structures and its interactions with electric, magnetic and strain fields.
\end{abstract}

\pacs{31.15.xh; 71.70.Ej; 76.30.Mi}

\maketitle

\section{Introduction}

The negatively charged nitrogen-vacancy (NV$^-$) center is a unique defect in diamond that has many promising applications in quantum metrology and quantum information processing (QIP). In particular, the ground state spin has been used in recent demonstrations of high precision magnetic \cite{mag1,mag2,mag3,mag4,mag5,mag6,mag7,mag8} and electric field \cite{efield} sensing, as well as spin-photon \cite{spinphoton} and spin-spin \cite{spinspin1,spinspin2,spinspin3,spinspin4,spinspin5} entanglement. The NV$^-$ center has also been employed to explore the developing decoherence based sensing techniques. \cite{decoherence1,decoherence2,decoherence3,decoherence4} Each of these demonstrations exploit the interaction of the spin with some configuration of electric, magnetic and strain fields and the center's remarkable capability of optical spin-polarization and readout. \cite{readout1,readout2} The demonstration of electric field sensing, \cite{efield} which required the precise control of the spin using magnetic fields and the intricate modeling of the spin's interaction with electric, magnetic and strain fields, highlighted the NV$^-$ center as a universal field sensor at the nanoscale as well as emphasized the requirement for a more detailed theoretical understanding of this important spin in diamond. The development of such an understanding will enable more precise control and modeling of the spin in its current applications and also provide the necessary insight to explore its future applications.

\begin{figure}[hbtp]
\begin{center}
\includegraphics[width=1\columnwidth] {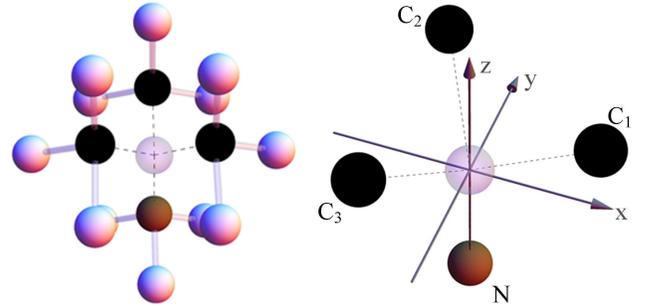}
\caption{(color online) Schematics of the nitrogen-vacancy center and lattice depicting the vacancy (transparent), the nearest-neighbor carbon atoms to the vacancy (black), the substitutional nitrogen atom (brown), and the next-to-nearest carbon neighbors to the vacancy (white). The adopted coordinate system and carbon labels are depicted in the right schematic.}
\label{fig:center}
\end{center}
\end{figure}

\begin{figure}[hbtp]
\begin{center}
\includegraphics[width=0.9\columnwidth] {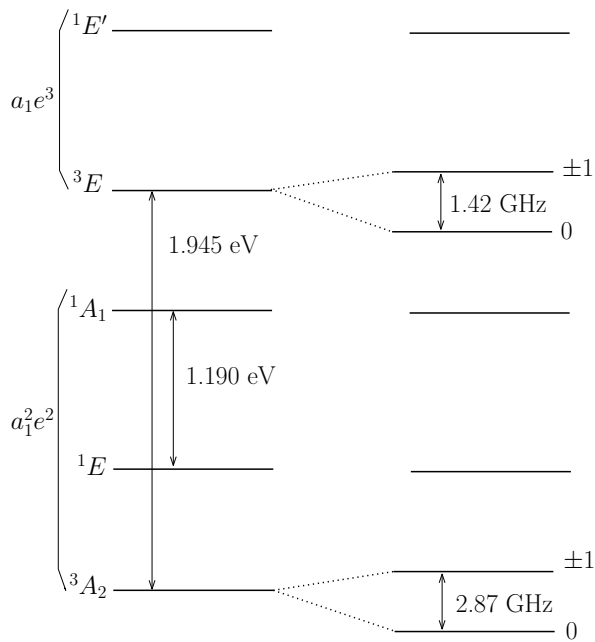}
\caption{The electronic orbital structure (left) and fine structure (right) at ambient temperatures formed from the ground $a_1^2e^2$ and first excited $a_1e^3$ molecular orbital configurations. The observed optical zero phonon line (1.945 eV) \cite{dupreez} and infrared zero phonon line (1.190 eV) \cite{infrared} transitions are depicted as solid arrows in the orbital structure.}
\label{fig:electronicstructure}
\end{center}
\end{figure}

The NV$^-$ center is a point defect of $C_{3v}$ symmetry in diamond consisting of a substitutional nitrogen atom adjacent to a carbon vacancy (refer to Fig. \ref{fig:center}). The center's electronic structure is summarized in Fig. \ref{fig:electronicstructure}. It consists of a $^3A_2$ ground triplet state, an optical $^3E$ excited triplet and several dark singlet states. \cite{njp} The fine structure of the $^3E$ excited triplet is highly dependent on temperature \cite{averaging} and crystal strain, \cite{excitedstatestrain} whereas the fine structure of the $^3A_2$ ground triplet state is observed to be only weakly dependent on temperature \cite{acostatemp} and crystal strain with a single zero-field splitting of $D_{gs}\sim 2.87$ GHz between the $m_s = 0$ and $m_s = \pm1$ spin sub-levels. At ambient temperatures, the fine structure of the excited triplet state replicates the ground triplet state with a single zero-field splitting of $D_{es} \sim 1.42$ GHz \cite{excitedstatezeeman,excitedstatezeeman2} independent of crystal strain due to the dynamic Jahn-Teller effect. \cite{averaging} Zeeman and Stark splittings have been observed in the fine structures of both triplet states, \cite{excitedstatezeeman,ground,tamarat,vanoort} although the Stark effect in the ground triplet state is several orders of magnitude smaller than in the excited triplet state. \cite{tamarat,vanoort}

Electron spin resonance (ESR) \cite{xing} and \textit{ab initio} studies \cite{ab initio1,ab initio2,ab initio3,ab initio4,ab initio5,ab initio6,ab initio7,ab initio8,galihyperfine} have confirmed that the electronic states of the center are highly localized to the vacancy and its nearest neighbors. The localization of the center's electronic states supports the application of a molecular model, in which the center's states are approximated by configurations of molecular orbitals (MOs).  The molecular model has been successfully applied to describe the effects of electric, magnetic and strain fields on the fine structure of the excited triplet state, \cite{ excitedstatestrain,excitedstatezeeman, tamarat} however the model has not yet been applied to describe the effects of the fields on the ground triplet state. This has been due to an absence of the spin-orbit and spin-spin induced couplings of the center's electronic states, which have been obtained recently in Ref. \onlinecite{njp}. Consequently, until now, the measurements of the effects of the fields on the ground state spin have been interpreted using the effective spin-Hamiltonian formulism of ESR. \cite{vanoort,xing,loubser,gfactor} Whilst the spin-Hamiltonian formulism has provided a practical model for the implementation of the center's applications to date, it does not facilitate the correlation of the properties of the ground state spin to the other properties of the center nor the theoretical prediction of more intricate properties of the spin. Consequently, the application of the molecular model to describe the effects of electric, magnetic and strain fields on the ground state spin is expected to provide the required detailed theoretical understanding of the spin.

In this article, the well established molecular model \cite{njp,loubser,lenef,mansonmodel} of the NV$^-$ center will be applied in order to fully develop the theory of the ground state spin. The fine and hyperfine structure and their corresponding eigenstates will be constructed prior to examining the effect of electric, magnetic and strain fields on each. By using the matrix representations derived in the recently published electronic solution, \cite{njp} explicit expressions in terms of the center's MOs will be derived in this work  for the hyperfine interaction with the $^{14}\mathrm{N}$ nucleus, the components of the electron g-factor tensor, and the Stark and strain interactions. The derivation of these expressions enables the rigorous definition of the spin-Hamiltonian of the ground state spin and the correlation of the accurately measured parameters of the spin to the other observed properties of the center. The expressions will also assist future \textit{ab initio} studies to independently calculate the properties of the ground state spin.

\section{Electronic fine structure and interactions with electric, magnetic and strain fields}

By adopting the adiabatic approximation and considering the nuclei of the crystal to be fixed at their equilibrium coordinates $\vec{R}_0$ corresponding to the ground electronic state, the electronic Hamiltonian $\hat{H}_e$ of the NV$^-$ center can be defined as
\begin{eqnarray}
\hat{H}_{e} & = & \sum_i\hat{T}_i+\hat{V}_{Ne}(\vec{r}_i,\vec{R}_0)+\hat{V}_{so}(\mathbf{x}_i,\vec{R}_0)+\hat{V}_{hf}(\mathbf{x}_i,\mathbf{X}_0) \nonumber \\
&&+\sum_{i>j}\hat{V}_{ee}(\mathbf{x}_i,\mathbf{x}_j)+\hat{V}_{ss}(\mathbf{x}_i,\mathbf{x}_j)
\label{eq:NVhamiltonian}
\end{eqnarray}
where $\mathbf{x}_i = (\vec{r}_i,\vec{s}_i)$ denotes the collective spatial and spin coordinates of the $i^{th}$ electron of the center, $\mathbf{X}_0 = (\vec{R}_0,\vec{I})$ denotes the collective equilibrium spatial and spin coordinates of the crystal nuclei, $\hat{T}_i$ is the kinetic energy of the $i^{th}$ electron, $\hat{V}_{Ne}$ is the effective Coulomb potential of the interaction of the nuclei and lattice electrons with the electrons of the center, $\hat{V}_{so}$ is the electronic spin-orbit potential, $\hat{V}_{hf}$ is the hyperfine potential of the interactions between the crystal nuclei and the electrons of the center, $\hat{V}_{ee}$ is the Coulomb repulsion potential of the electrons of the center and $\hat{V}_{ss}$ is the electronic spin-spin potential.

\textit{Ab initio} studies \cite{ab initio1,ab initio2,ab initio3,ab initio4,ab initio5,ab initio6,ab initio7,ab initio8,galihyperfine} have confirmed the presence of three MOs ($a_1$, $e_x$, $e_y$) in the band gap of diamond and the center's observable electronic structure has been shown to consist of the ground $a_1^2e^2$ and first excited $a_1e^3$ MO configurations formed from the occupation of the MOs by four electrons. \cite{njp} Note that the other two electrons of the six electrons associated with the center occupy delocalized $A_1$ symmetric MOs within the diamond valence band and do not influence the observable properties of the center. The electronic states can be constructed by firstly defining orbital states with well defined $C_{3v}$ orbital symmetry formed from products of the MOs and defining spin states with well defined $C_{3v}$ spin symmetry. \cite{njp} Secondly, electronic states $\Phi_{n,j,k}^{so}$ that transform as specific rows $k$ of irreducible representations $j$ of the $C_{3v}$ group in spin-orbit space are formed from linear combinations of the orbital and spin state products. \cite{njp} Note that the quantum number $n$ denotes the fine structure level of the electronic state. For example, using the irreducible representations contained in Ref. \onlinecite{lenef}, the electronic states of the ground triplet are
\begin{eqnarray}
\Phi_{1,A_1}^{so} = & \Phi_{A_2}S_{A_2} = \frac{1}{\sqrt{2}}(\ket{a_1\bar{a}_1e_x\bar{e}_y}+\ket{a_1\bar{a}_1\bar{e}_xe_y}) \nonumber \\
\Phi_{2,E,x}^{so} = & -\Phi_{A_2}S_{E,y} = \frac{1}{\sqrt{2}}(\ket{a_1\bar{a}_1e_xe_y}-\ket{a_1\bar{a}_1\bar{e}_x\bar{e}_y}) \nonumber \\
\Phi_{2,E,y}^{so} = & \Phi_{A_2}S_{E,x} = \frac{-i}{\sqrt{2}}(\ket{a_1\bar{a}_1e_xe_y}+\ket{a_1\bar{a}_1\bar{e}_x\bar{e}_y})
\end{eqnarray}
where $\Phi_{A_2} = \frac{1}{\sqrt{2}}(a_1a_1e_xe_y-a_1a_1e_ye_x)$ is the $A_2$ orbital state of the ground triplet, the kets on the right hand side denote Slater determinants (overbar denoting down-spin), and the symmetrised $S=1$ spin states in terms of the $S_z$ eigenstates $\{\ket{S,m_s}\}$ are $S_{A_2} = \ket{1,0}$, $S_{E,x} = \frac{-i}{\sqrt{2}}(\ket{1,1}+\ket{1,-1})$ and $S_{E,y} = \frac{-1}{\sqrt{2}}(\ket{1,1}-\ket{1,-1})$.

The majority of the spin-orbit states $\Phi_{n,j,k}^{so}$ are eigenstates of the orbital components of the electronic Hamiltonian $\hat{H}_o = \sum_i\hat{T}_i+\hat{V}_{Ne}(\vec{r}_i,\vec{R}_0)+\sum_{i>j}\hat{V}_{ee}(\mathbf{x}_i,\mathbf{x}_j)$ with orbital energies denoted by $E_{J;S}$ (where to remain consistent with Ref. \onlinecite{njp}, $J$ denotes the irreducible representation of the orbital state and $S$ denotes the total spin of the spin states used to form  $\Phi_{n,j,k}^{so}$ and the orbital energy of the ground triplet is defined as $E_{A_2;1}=0$). Only the spin-orbit states $(\Phi_{3,E,x}^{so},\Phi_{3,E,y}^{so},\Phi_{9,E,x}^{so},\Phi_{9,E,y}^{so})$ associated with the $^1E$ and $^1E'$ singlets are not eigenstates of $\hat{H}_o$ and are mixed by the Coulomb coupling coefficient $\kappa$ (refer to Ref. \onlinecite{njp} for further details). The energies $E_{J;S}$ including the effects of the Coulomb repulsion of the $E$ singlets form the orbital structure of the center depicted in Fig. \ref{fig:electronicstructure}.

The electronic spin-orbit and spin-spin potentials can be treated as first-order perturbations to $\hat{H}_o$ using the orbital energies $E_{J;S}$ and the spin-orbit states $\Phi_{n,j,k}^{so}$ (accounting for Coulomb coupling) as the zero-order energies and states of the perturbation expansion. The perturbed energies correct to first-order $E_n$ have been shown to be consistent with the observed fine structure of the center depicted in Fig. \ref{fig:electronicstructure}. \cite{excitedstatestrain} It is found that the fine structure of the ground triplet state is governed by electronic spin-spin interaction which splits the $m_s = 0$ and $m_s = \pm1$ spin sub-levels such that $E_2-E_1 = D_{gs}\sim2.87$ GHz, where
\begin{eqnarray}
D_{gs} & = & \frac{3\mu_0g_e^2\mu_B^2}{8\pi}\bra{e_x(\vec{r}_1)e_y(\vec{r}_2)}\frac{1-3z_{12}^2/|\vec{r}_{12}|^2}{|\vec{r}_{12}|^3} \nonumber \\
&& (\ket{e_x(\vec{r}_1)e_y(\vec{r}_2)}-\ket{e_y(\vec{r}_1)e_x(\vec{r}_2)}),
\end{eqnarray}
$\mu_0$ is the vacuum permeability, $g_e=2.0023$ is the free electron g-factor, $\mu_B$ is the Bohr magneton, $\vec{r}_i = x_i\vec{x}+y_i\vec{y}+z_i\vec{z}$ ($\vec{x}$,$\vec{y}$,$\vec{z}$ being unit coordinate vectors), $\vec{r}_{12} = \vec{r}_2-\vec{r}_1$, and $z_{12} = z_2-z_1$. As obtained in Ref. \onlinecite{njp}, the first-order corrected spin-orbit states $\Phi_{n,j,k}^{so\prime}$ have the general form
\begin{eqnarray}
\Phi_{n,j,k}^{so\prime} = N_n\left(\Phi_{n,j,k}^{so}+\sum_{m\neq n} s_{n,m}\Phi_{m,j,k}^{so}\right)
\end{eqnarray}
where $N_n$ are normalization constants and $s_{n,m}$ are the first-order spin coupling coefficients. Using the results of Ref. \onlinecite{njp}, the first-order spin-orbit states of the ground triplet are
\begin{eqnarray}
\Phi_{1,A_1}^{so\prime} & = & \Phi_{1,A_1}^{so}+s_{1,4}\Phi_{4,A_1}^{so}+s_{1,8}\Phi_{8,A_1}^{so} \nonumber \\
\Phi_{2,E,k}^{so\prime}  & = & \Phi_{2,E,k}^{so}+s_{2,3}\Phi_{3,E,k}^{so}+s_{2,5}\Phi_{5,E,k}^{so} \nonumber \\
&&+s_{2,6}\Phi_{6,E,k}^{so}+s_{2,9}\Phi_{9,E,k}^{so} \label{eq:fosostates}
\end{eqnarray}
where $k = x,y$ and the spin coupling coefficients of the ground triplet are contained in table \ref{tab:spincoefficients} and are functions of the orbital energies $E_{J;S}$, the axial $\lambda_\parallel = 5.3$ GHz \cite{excitedstatestrain} and non-axial $\lambda_\perp\sim$ GHz spin-orbit parameters, the spin-spin parameters $D_{1,E,1}\sim$ MHz and $D_{1,E,2}\sim$ MHz, the Coulomb coupling coefficient $\kappa$, and the spin-spin coupling coefficient of the excited triplet $\eta = 0.053$. \cite{njp}

\begin{table}
\caption{\label{tab:spincoefficients} The spin coupling coefficients of the ground triplet correct to first-order in spin-orbit and spin-spin interactions. The Coulomb coupling coefficient $\kappa$, the spin-orbit parameters ($\lambda_\parallel$, $\lambda_\perp$), the spin-spin parameters ($D_{1,E,1}$,$D_{1,E,2}$), and the spin-spin coupling coefficient of the excited triplet $\eta$ are as defined in Ref. \onlinecite{njp}. Note that $N_\kappa = (1+|\kappa|^2)^{-1/2}$.}
\begin{ruledtabular}
\begin{tabular}{l}
$s_{1,4}$ = $-2i\frac{\lambda_\parallel}{E_{A_1;0}}$ \\
$s_{1,8}$ = $-\sqrt{2}\frac{\lambda_\perp+D_{1,E,2}}{E_{E;1}}$ \\
$s_{2,3}$ = $iN_\kappa\kappa\frac{\lambda_\perp}{E_{E;0}}$ \\
$s_{2,5}$ = $-\sqrt{2}\frac{D_{1,E,1}}{E_{E;1}}-\eta\frac{\lambda_\perp-D_{1,E,2}}{E_{E;1}}$ \\
$s_{2,6}$ = $\frac{\lambda_\perp-D_{1,E,2}}{E_{E;1}}-\sqrt{2}\eta\frac{D_{1,E,1}}{E_{E;1}}$ \\
$s_{2,9}$ = $-iN_\kappa\frac{\lambda_\perp}{E_{E^\prime;0}}$ \\
\end{tabular}
\end{ruledtabular}
\end{table}

Note that recent strain measurements of the infrared zero phonon line (ZPL) \cite{irstrain} have indicated that the Coulomb coupling coefficient is significant $\kappa\sim0.3$ and therefore must be retained to second-order in the spin coupling coefficients. Furthermore, it should be noted that since the spin-orbit and spin-spin parameters are expected to be of the order of $\sim$$10^{-9}$-$10^{-6}$ eV (MHz-GHz) and the orbital energies are expected to be of the order of $\sim$ $10^{-2}-10^1$ eV, the spin coupling coefficients are expected to be of the order of $10^{-4}-10^{-10}$. The relative magnitudes of the different coefficients will become important in determining the leading order terms that contribute to the interactions of the ground state spin with electric, magnetic and strain fields.

Given (\ref{eq:fosostates}) and the zero-order orbital and spin operator matrix representations contained in Ref. \onlinecite{njp}, matrix representations of the ground triplet using the basis of first-order corrected spin-orbit states  $\{\Phi_{1,A_1}^{so\prime},\Phi_{2,E,x}^{so\prime},\Phi_{2,E,y}^{so\prime}\}$ can be constructed for general one-electron orbital tensor operators $\hat{O}_{p,q} = \sum_i \hat{O}_{p,q}(\vec{r}_i)$, which transform as the row $q$ of the  irreducible representation $p$ of the $C_{3v}$ group, and the total spin operator $\vec{S}=\sum_i\vec{s}_i$ (see table \ref{tab:spinandorbitaloperator}). The matrix representations are expressed in their most simplified form in terms of one-electron reduced matrix elements and the center's MOs. The one-electron matrix elements and the associated reduced matrix elements are related by the Wigner-Eckart theorem \cite{cornwell}
\begin{eqnarray}
\me{\phi_{f,g}(\vec{r}_1)}{\hat{O}_{p,q}(\vec{r}_1)}{\phi_{j,k}(\vec{r}_1)} = \cg{j}{k}{p}{q}{f}{g}^\ast\rdm{\phi_f}{\hat{O}_p}{\phi_j}\label{eq:wignereckart}\notag\\
\end{eqnarray}
where $\phi_{j,k}$ and $\phi_{f,g}$ are MOs of symmetry $(j,k)$ and $(f,g)$ respectively, and $( \ |)$ are the Clebsch-Gordon coefficients defined in Ref. \onlinecite{lenef}. The detailed model of the interactions of the ground state spin with electric, magnetic and strain fields can thus be developed by applying the matrix representations of table \ref{tab:spinandorbitaloperator} to each interaction in turn. As a result, the interactions will be expressed in their most simplified form in terms of the spin coupling coefficients, the MOs, and the reduced matrix elements.

\begin{table*}
\caption{\label{tab:spinandorbitaloperator} Matrix representations of the components of the total spin operator $\vec{S}=\sum_i\vec{s}_i$ and the orbital tensor operators $\hat{O}_{p,q}=\sum_i\hat{O}_{p,q}(\vec{r}_i)$ of different symmetry $(p,q)$ correct to first-order in the spin coupling coefficients in the basis of the corrected spin-orbit states $\{\Phi_{1,A_1}^{so\prime},\Phi_{2,E,x}^{so\prime},\Phi_{2,E,y}^{so\prime}\}$ of the ground triplet. The orbital parameters in terms of reduced matrix elements of the center's molecular orbitals are $o_{a,A_1} = 2(\rdm{a_1}{\hat{O}_{A_1}}{a_1}+\rdm{e}{\hat{O}_{A_1}}{e})$ and $o_{a,E} = \rdm{a_1}{\hat{O}_{A_1}}{e})$.}
\begin{ruledtabular}
\begin{tabular}{c}
$S_x = \left(\begin{array}{ccc}
0 & 0 & -i\hbar \\
0 & 0 & 0 \\
i\hbar & 0 & 0 \\
\end{array}\right)$
$S_y = \left(\begin{array}{ccc}
0 & i\hbar & 0 \\
-i\hbar & 0 & 0 \\
0 & 0 & 0 \\
\end{array}\right)$
$S_z = \left(\begin{array}{ccc}
0 & 0 & 0 \\
0 & 0 & -i\hbar\\
0 & i\hbar & 0 \\
\end{array}\right)$ \\
$\hat{O}_{A_1} = \left(\begin{array}{ccc}
o_{a,A_1} & 0 & 0 \\
0 & o_{a,A_1} & 0 \\
0 & 0 & o_{a,A_1} \\
\end{array}\right)$ \ \
$\hat{O}_{A_2} = \left(\begin{array}{ccc}
0 & 0 & 0 \\
0 & 0 & 0 \\
0 & 0 & 0 \\
\end{array}\right)$ \\
$\hat{O}_{E,x} = \left(\begin{array}{ccc}
0 & s_{2,6}o_{a,E}+\frac{s_{1,8}}{\sqrt{2}}o_{a,E}^\ast & 0 \\
s_{2,6}o_{a,E}^\ast+\frac{s_{1,8}}{\sqrt{2}}o_{a,E} & -\frac{1}{\sqrt{2}}s_{2,5}(o_{a,E}+o_{a,E}^\ast) & 0 \\
0 & 0 & \frac{1}{\sqrt{2}}s_{2,5}(o_{a,E}+o_{a,E}^\ast) \\
\end{array}\right)$ \\
$\hat{O}_{E,y} = \left(\begin{array}{ccc}
0 & 0 & s_{2,6}o_{a,E}+\frac{s_{1,8}}{\sqrt{2}}o_{a,E}^\ast \\
0 & 0 & \frac{1}{\sqrt{2}}s_{2,5}(o_{a,E}+o_{a,E}^\ast) \\
s_{2,6}o_{a,E}^\ast+\frac{s_{1,8}}{\sqrt{2}}o_{a,E} & \frac{1}{\sqrt{2}}s_{2,5}(o_{a,E}+o_{a,E}^\ast) & 0 \\
\end{array}\right)$ \\
\end{tabular}
\end{ruledtabular}
\end{table*}

\subsection{Interactions with magnetic fields}

Defining $\vec{B}$ to be the applied magnetic field that is assumed to be approximately constant over the dimensions of the NV$^-$ center, the interaction of the center's electrons with the magnetic field is described by the potential \cite{zeemaninteraction}
\begin{eqnarray}
\hat{V}_\mathrm{mag} & = & \frac{\mu_B}{\hbar}\sum_i(\vec{l}_i+g_e\vec{s}_i)\cdot\vec{B} \nonumber \\
&&+ \frac{1}{2m_ec^2}(\vec{s}_i\times\vec{\nabla}\hat{V}_{Ne}(\vec{r}_i))\cdot(\vec{B}\times\vec{r}_i)\label{eq:magneticinteraction}
\end{eqnarray}
where $\vec{l}=l_{x}\vec{x}+l_{y}\vec{y}+l_{z}\vec{z}=\vec{r}\times\vec{p}$ is the electron orbital magnetic moment, $\vec{p}$ is the electron momentum, $\hbar$ is the reduced Planck constant, $m_e$ is the mass of an electron and $c$ is the speed of light. Note that the origin of the coordinate system is defined to be at the center of the NV$^-$ defect, in the vicinity of the vacancy (as depicted in Fig. \ref{fig:center}). Additionally, note that the term $\sum_i\frac{e^2}{8m_e}(\vec{B}\times\vec{r}_i)^2$ (where $e$ is the electronic charge) quadratic in the magnetic field \cite{zeemaninteraction} has been neglected in the above definition since it does not induce a relative shift of the fine structure levels  or couple the electronic states of the ground triplet at first-order in the spin coupling coefficients. The second term in (\ref{eq:magneticinteraction}) arises from relativistic corrections to the non-relativistic first term \cite{zeemaninteraction} and can be written in the more explicit form
\begin{eqnarray}
\frac{1}{2m_ec^2}(\vec{s}\times\vec{\nabla}\hat{V}_{Ne})\cdot(\vec{B}\times\vec{r})= \frac{1}{2m_ec^2}\vec{s}\cdot\bar{G}\cdot\vec{B} \label{eq:magrelterm}
\end{eqnarray}
where the orbital operator $\bar{G}$ is the matrix
\begin{eqnarray}
\left(\begin{array}{ccc}
\frac{\partial\hat{V}_{Ne}}{\partial y}y+\frac{\partial \hat{V}_{Ne}}{\partial z}z & -\frac{\partial\hat{V}_{Ne}}{\partial y}x & -\frac{\partial\hat{V}_{Ne}}{\partial z}x \\
-\frac{\partial\hat{V}_{Ne}}{\partial x}y & \frac{\partial\hat{V}_{Ne}}{\partial x}x+\frac{\partial\hat{V}_{Ne}}{\partial z}z & -\frac{\partial\hat{V}_{Ne}}{\partial z}y \\
-\frac{\partial\hat{V}_{Ne}}{\partial x}z & -\frac{\partial\hat{V}_{Ne}}{\partial y}z & \frac{\partial\hat{V}_{Ne}}{\partial x}x+\frac{\partial\hat{V}_{Ne}}{\partial y}y \\
\end{array}\right)\nonumber
\end{eqnarray}

The tensor components of the orbital operators contained in (\ref{eq:magneticinteraction}) are
\begin{eqnarray}
&& l_{E,x} = l_y, \ l_{E,y} = -l_x, \ l_{A_2} = l_z \nonumber \\
&& G_{1,A_1} = \frac{\partial\hat{V}_{Ne}}{\partial x}x+\frac{\partial\hat{V}_{Ne}}{\partial y}y, \ G_{2,A_1} = \frac{\partial\hat{V}_{Ne}}{\partial z}z, \nonumber \\
&& G_{1,E,x} = \frac{\partial\hat{V}_{Ne}}{\partial x}x-\frac{\partial\hat{V}_{Ne}}{\partial y}y, \ G_{2,E,x} = \frac{\partial\hat{V}_{Ne}}{\partial z}x, \nonumber \\
&& G_{3,E,x} = \frac{\partial\hat{V}_{Ne}}{\partial x}z, \  G_{1,E,y} = -\frac{\partial\hat{V}_{Ne}}{\partial x}y-\frac{\partial\hat{V}_{Ne}}{\partial y}x, \nonumber \\
&& \ G_{2,E,y} = \frac{\partial\hat{V}_{Ne}}{\partial z}y, \ G_{3,E,y} = \frac{\partial\hat{V}_{Ne}}{\partial y}z
\end{eqnarray}
Hence, using the general matrix representations of table \ref{tab:spinandorbitaloperator}, the matrix representation of the magnetic interaction (\ref{eq:magneticinteraction}) with the ground triplet can be obtained correct to first-order in the spin coupling coefficients
\begin{eqnarray}
V_\mathrm{mag} = \mu_B\left(\begin{array}{ccc}
0 & ig_\perp B_y & -ig_\perp B_x \\
-ig_\perp B_y & 0 & -ig_\parallel B_z \\
ig_\perp B_x & ig_\parallel B_z & 0 \\
\end{array}\right)\label{eq:magmatrixrep}
\end{eqnarray}
where
\begin{eqnarray}
&& g_\perp =  g_e+(s_{2,6}-\frac{s_{1,8}}{\sqrt{2}})l_{a,E}+\frac{1}{2}g_{1,A_1}+g_{2,A_1}, \nonumber \\
&& g_\parallel =  g_e+g_{1,A_1} \label{eq:gfactors}
\end{eqnarray}
and $l_{a,E} = -i\rdm{a_1}{l_E}{e}/\sqrt{2}\hbar$, $g_{1,A_1} = \rdm{e}{G_{1,A_1}}{e}/2m_ec^2$ and $g_{2,A_1} = \rdm{e}{G_{2,A_1}}{e}/2m_ec^2$. Given that the relativistic term (\ref{eq:magrelterm}) is much smaller than the non-relativistic term, only contributions of the relativistic term that are zero-order in the spin coupling coefficients have been included.

Comparing (\ref{eq:magmatrixrep}) with the matrix representations of the total spin operators in  table \ref{tab:spinandorbitaloperator}, it can be seen that the magnetic interaction can be written in the spin-Hamiltonian form $\hat{V}_{mag} = \frac{\mu_B}{\hbar}\vec{S}\cdot\bar{g}\cdot\vec{B}$, where the effective g-factor tensor $\bar{g}$ is defined as
\begin{eqnarray}
\bar{g} = \left(\begin{array}{ccc}
g_\perp & 0 & 0 \\
0 & g_\perp & 0 \\
0 & 0 & g_\parallel \\
\end{array}\right) \label{eq:gtensor}
\end{eqnarray}
The parameters $g_\perp$ and $g_\parallel$ may then be identified as the non-axial and axial effective g-factor components respectively. The g-factor components have been measured by several ESR studies \cite{xing,loubser,gfactor} and the observed values are contained in table \ref{tab:ESRparameters}. The measurements conclusively show that both $g_\parallel$ and $g_\perp$ are shifted by $+6\pm1\times10^{-4}$ from the free electron g-factor $g_e = 2.0023$ and one study \cite{gfactor} observed a small anisotropy of $g_\perp-g_\parallel = 2\times10^{-4}$. From (\ref{eq:gfactors}) it is clear that there is no orbital magnetic moment contribution to $g_\parallel$ at first-order in the spin coupling coefficients and due to the expected orders of magnitudes of the coefficients, any second-order contribution would be much smaller than the observed shifts of $g_\parallel$ and $g_\perp$. Thus, only the relativistic term (\ref{eq:magrelterm}) which shifts $g_\parallel$ from $g_e$ by the addition of $g_{1,A_1}$ and also shifts $g_\perp$ by the addition of $\frac{1}{2}g_{1,A_1}+g_{2,A_1}$, can explain the observed shifts. The different contributions of the relativistic term to $g_\perp$ and $g_\parallel$ can also explain the small anisotropy, however if the orbital magnetic moment contribution to $g_\perp$ is large enough, it may also contribute at the same order as the relativistic term. The leading order term of the orbital magnetic moment contribution to $g_\perp$ is
\begin{eqnarray}
(s_{2,6}-\frac{s_{1,8}}{\sqrt{2}})l_{a,E}\approx 2\frac{\lambda_\perp}{E_{E;1}}l_{a,E}
\end{eqnarray}
due to the expectation that $\lambda_\perp \gg \eta D_{1,E,2}$. Through the observation of the magnetic circular dichroism (MCD) of the optical ZPL, \cite{averaging} the orbital magnetic moment parameter will be of the order $l_{a,E}\sim10^{-1}$. Consequently, the orbital magnetic moment will only contribute significantly to the g-factor anisotropy if $2\frac{\lambda_\perp}{E_{E;1}}\sim 10^{-3}$, which given $E_{E;1}\approx 2.180$ eV, \cite{njp} implies that $\lambda_\perp$ would have to be of the order of 1 meV $\sim 1$ THz, a dramatic difference from the axial spin-orbit parameter $\lambda_\parallel = 5.3$ GHz. Such a dramatic difference is improbable given the NV$^-$ center's small departure from the higher $T_d$ symmetry, in which the axial and non-axial spin-orbit parameters are equal. Therefore, it appears unlikely that the orbital magnetic moment contributes significantly to $\bar{g}$.

\begin{table*}
\caption{\label{tab:ESRparameters} Experimental measurements and \textit{ab initio} calculations of the effective g-factor and hyperfine parameters of the NV$^-$ center tabulated by reference.}
\begin{ruledtabular}
\begin{tabular}{llllll}
Ref. & $g_\perp$ & $g_\parallel$ & $A_\perp/h$ (MHz) & $A_\parallel/h$ (MHz) & $P/h$ (MHz) \\
\hline
Loubser \cite{loubser} (Exp) & 2.0028(3) & 2.0028(3) & - & $(\pm)2.32\pm0.01$ & - \\
He \cite{xing} (Exp) & 2.0028(3) & 2.0028(3) & $(+)2.10\pm0.10$ & $(+)2.30\pm0.02$ & $(-)5.04\pm0.05$ \\
Felton \cite{gfactor} (Exp) & 2.0031(2) & 2.0029(2) & $(-)2.70(7)$ & $(-)2.14(7)$ & $(-)5.01(6)$ \\
Steiner \cite{steiner} (Exp) & - & - & - & $-2.166\pm0.01$ & $-4.945\pm0.01$ \\
Smeltzer \cite{smeltzer} (Exp) & - & - & - & $-2.162(2)$ & $-4.945(5)$ \\
Gali \cite{galihyperfine} (\textit{Ab initio}) & - & - & $(-)1.7$ & $(-)1.7$ & - \\
\end{tabular}
\end{ruledtabular}
\end{table*}

\subsection{Interactions with electric fields}

Defining $\vec{E}$ to be the applied electric field that is assumed to be approximately constant over the dimensions of the NV$^-$ center, the interaction of the center's electrons with the electric field is described by the potential \cite{MQM}
\begin{eqnarray}
\hat{V}_\mathrm{el} = \sum_i\vec{d}_i\cdot\vec{E}\label{eq:electricinteraction}
\end{eqnarray}
where $\vec{d} = e\vec{r}$ is the electron electric dipole moment. The tensor components of $\vec{d}$ are simply $\vec{d} = d_{E,x}\vec{x}+d_{E,y}\vec{y}+d_{A_1}\vec{z}$ and the ground triplet matrix representation of $\hat{V}_{el}$ correct to first-order in the spin coupling coefficients is
\begin{eqnarray}
V_\mathrm{el} = \left(\begin{array}{ccc}
d_{a,A_1}E_z & d_\perp^\prime E_x & d_\perp^\prime E_y \\
d_\perp^\prime E_x & d_{a,A_1}E_z+d_\perp E_x  & -d_\perp E_y \\
d_\perp^\prime E_y & -d_\perp E_y & d_{a,A_1}E_z-d_\perp E_x \\
\end{array}\right) \nonumber \\
\end{eqnarray}
where $d_{a,A_1} = 2(\rdm{a_1}{d_{A_1}}{a_1}+\rdm{e}{d_{A_1}}{e})$, $d_\perp = -2s_{2,5}d_{a,E}$, $d_\perp^\prime = (s_{2,6}+\frac{s_{1,8}}{\sqrt{2}})d_{a,E}$ and $d_{a,E} = \frac{1}{\sqrt{2}}\rdm{a_1}{d_E}{e}$.

The leading order terms of the two non-axial dipole parameters are
\begin{eqnarray}
d_\perp \approx 2\frac{\eta\lambda_\perp}{E_{E;1}}d_{a,E}, \ \ d_\perp^\prime \approx -2\frac{D_{1,E,2}}{E_{E;1}}d_{a,E}
\end{eqnarray}
which due to the potentially similar orders of magnitude of the numerators in each, suggests that the parameters are potentially of the same magnitude. However, since $d_\perp^\prime$ couples electronic states separated in energy by $D_{gs}$, it can be ignored for static electric fields that satisfy $d_\perp^\prime E_\perp \ll D_{gs}$, where $E_\perp = \sqrt{E_x^2+E_y^2}$ is the non-axial electric field strength. This conclusion is in agreement with observation, \cite{efield,vanoort} where small linear Stark splittings of the $m_s = \pm1$ fine structure levels have been shown to agree with $d_\perp^\prime \approx 0$ and $d_\perp/h = 17\pm3$ Hz cm/V. \cite{vanoort}

Since correct to first-order in the spin coupling coefficients, the term $d_{a,A_1}E_z$ is common to each of the diagonal matrix elements,  it appears that an axial electric field does not induce a relative shift of the fine structure levels of the ground triplet. However, this conclusion is in conflict with observation, \cite{vanoort} where a very small linear shift of the zero-field splitting between the $m_s = 0$ and $m_s = \pm1$ fine structure levels was observed and could be described only by a difference of $d_\parallel/h = 0.35\pm0.02$ Hz cm/V in the axial dipole parameters of the $m_s = 0$ and $m_s = \pm1$ spin-orbit states. Such a difference occurs at second-order in the spin coupling coefficients, where the matrix representation of the interaction of the ground triplet with an axial electric field becomes
\begin{eqnarray}
d_{A_1}E_z = \left(\begin{array}{ccc}
d_{a,A_1}E_z & 0 & 0 \\
0 & (d_{a,A_1}+d_\parallel)E_z & 0 \\
0 & 0 &  (d_{a,A_1}+d_\parallel)E_z  \\
\end{array}\right) \nonumber \\
\end{eqnarray}
where $d_\parallel = (s_{2,5}^2+s_{2,6}^2+s_{2,9}^2)d_{b,A_1}$ and $d_{b,A_1} = \rdm{e}{d_{A_1}}{e}-\rdm{a_1}{d_{A_1}}{a_1}$. Therefore, the effective matrix representation of the interaction of the ground triplet with a static electric field satisfying $d_\perp^\prime E_\perp \ll D_{gs}$ is
\begin{eqnarray}
V_\mathrm{el} = \left(\begin{array}{ccc}
0 & 0 & 0 \\
0 & d_\parallel E_z+d_\perp E_x  & -d_\perp E_y \\
0 & -d_\perp E_y & d_\parallel E_z-d_\perp E_x \\
\end{array}\right)
\end{eqnarray}
This effective representation can be expressed in the spin-Hamiltonian form $\hat{V}_{el} = \frac{1}{\hbar^2}d_\parallel E_z S_z^2-\frac{1}{\hbar^2}d_\perp E_x(S_x^2-S_y^2)+\frac{1}{\hbar^2}d_\perp E_y(S_xS_y+S_yS_x)$ used to describe the linear Stark effect present in $C_{3v}$ symmetric systems in ESR. \cite{mims}

The dipole reduced matrix element $d_{a,E}$ contained in $d_\perp$ is also responsible for the center's optical transition \cite{njp} and an estimate of its magnitude can be obtained from the center's observed radiative lifetime $T_R\approx 13$ ns \cite{lifetime} using\cite{stoneham}
\begin{eqnarray}
d_{a,E}/h = \left(\frac{6\pi\epsilon_0\hbar^4c^3}{\langle E_{O}^3\rangle n_DT_R}\right)^{\frac{1}{2}}/h
\end{eqnarray}
where $\langle E_O^3\rangle = \int_0^\infty F(E_O)E_O^3dE_O$ is the expectation value of the cube of the optical emission energy given the normalized vibrational sideband distribution $F(E_O)$,\cite{vibtransition} and $n_D = 2.418$ is the refractive index of diamond. As the vibrational sideband of the center's optical emission extends from approximately $1.4$ eV to the ZPL at $1.945$ eV, the estimate of the dipole reduced matrix element is bounded by $3.65\leq d_{a,E}/h\leq 5.98$ MHz cm/V (compare with, for example, the 5.41 MHz cm/V dipole moment of the $5s($$^2S_{1/2})\longleftrightarrow 5p($$^2P_{3/2})$ transition of $^{87}$Rb).\cite{Rblifetime}

The observed value of $d_\perp/h = 17\pm3$ Hz cm/V, $\cite{vanoort}$ the estimated range of $d_{a,E}$, the approximate expression for $d_\perp \approx 2\frac{\eta\lambda_\perp}{E_{E;1}}d_{a,E}$, and $\eta = 0.053$, \cite{njp} imply that $\frac{\lambda_\perp}{E_{E;1}}\sim 10^{-4}$ as expected, thereby supporting the assertion made in the previous subsection that the orbital magnetic moment does not contribute significantly to $\bar{g}$. The dipole reduced matrix element $d_{b,A_1}$ contained in $d_\parallel$ also contributes to the shift of the center's optical ZPL in the presence of an axial electric field. \cite{njp} However, since the optical transition involves a change in MO configuration and, thus a change in the nuclear equilibrium coordinates, \cite{njp} both the axial electric and nuclear dipole moments contribute to the shift of the center's optical ZPL. Consequently, without knowledge of the nuclear dipole moment, it is not possible to estimate $d_{b,A_1}$ given just measurements of the shift.

\subsection{Interactions with strain fields}

The interaction of the center's electrons with a crystal strain field can be approximately described by performing a Taylor series expansion of the electronic Hamiltonian $\hat{H}_e$ in terms of the displacements of the nuclear coordinates $\vec{R}$ from their ground state equilibrium coordinates $\vec{R}_0$ induced by the strain field and retaining only the linear terms of the expansion. Defining $Q_{u,p,q}$ to be the $u^{th}$ normal nuclear displacement coordinate of the crystal that transforms as the row $q$ of the irreducible representation $p$ of the $C_{3v}$ group, the strain potential is \cite{stoneham}
\begin{eqnarray}
\hat{V}_\mathrm{str} = \sum_i\sum_{u,p,q}\left.\frac{\partial\hat{V}_{Ne}(\vec{r}_i,\vec{R})}{\partial Q_{u,p,q}}\right|_{\vec{R}_0}\xi_{u,p,q}
\end{eqnarray}
where $\xi_{u,p,q}$ is the crystal strain along the displacement coordinate $Q_{u,p,q}$. Given that by definition $\partial\hat{V}_{Ne}(\vec{r}_i,\vec{R})/\partial Q_{u,p,q}|_{\vec{R}_0}$ is an orbital tensor operator of symmetry $(p,q)$, the ground triplet matrix representation of $\hat{V}_{str}$ is analogous to that of $\hat{V}_{el}$. Treating the non-axial terms correct to first-order and the axial terms correct to second-order in the spin coupling coefficients, the matrix representation of $\hat{V}_{str}$ is
\begin{eqnarray}
V_\mathrm{str} = \left(\begin{array}{ccc}
\zeta_z^\prime & \zeta_x^\prime & \zeta_y^\prime \\
\zeta_x^\prime & \zeta_z^\prime+\zeta_z+\zeta_x  & -\zeta_y \\
\zeta_y^\prime & -\zeta_y & \zeta_z^\prime+\zeta_z-\zeta_x \\
\end{array}\right)
\end{eqnarray}
where
\begin{eqnarray}
\zeta_z^\prime & = & \sum_u\zeta_{u,a,A_1} \xi_{u,A_1} \notag \\
\zeta_z & = & (s_{2,5}^2+s_{2,6}^2+s_{2,9}^2)\sum_u\zeta_{u,b,A_1} \xi_{u,A_1}\notag \\
\zeta_k^\prime & = & (s_{2,6}+\frac{s_{1,8}}{\sqrt{2}})\sum_u\zeta_{u,a,E}\xi_{u,E,k}\notag \\
\zeta_k & = & -2s_{2,5}\sum_u\zeta_{u,a,E}\xi_{u,E,k}\notag \\
\zeta_{u,a,A_1} & = & 2\rdm{a_1}{\left.\frac{\partial\hat{V}_{Ne}}{\partial Q_{u,A_1}}\right|_{\vec{R}_0}}{a_1}+2\rdm{e}{\left.\frac{\partial\hat{V}_{Ne}}{\partial Q_{u,A_1}}\right|_{\vec{R}_0}}{e}\notag \\
\zeta_{u,b,A_1} & = & \rdm{e}{\left.\frac{\partial\hat{V}_{Ne}}{\partial Q_{u,A_1}}\right|_{\vec{R}_0}}{e}-\rdm{a_1}{\left.\frac{\partial\hat{V}_{Ne}}{\partial Q_{u,A_1}}\right|_{\vec{R}_0}}{a_1}\notag \\
\zeta_{u,a,E} & = & \frac{1}{\sqrt{2}}\rdm{a_1}{\left.\frac{\partial\hat{V}_{Ne}}{\partial Q_{u,E}}\right|_{\vec{R}_0}}{e}
\end{eqnarray}
and $k$ = $x$, $y$. Note that torsional strain components that have displacement coordinates that transform as $A_2$ have been ignored.

Similar to $\hat{V}_{el}$, the diagonal contributions $\zeta_z^\prime$ do not shift the fine structure levels of the ground triplet with respect to each other and thus can be effectively ignored. Likewise, for non-axial strains that satisfy $\zeta_\perp^\prime \ll D_{gs}$ (where $\zeta_\perp^\prime = \sqrt{\zeta_x^{\prime2}+\zeta_y^{\prime2}}$), $\zeta_x^\prime$ and $\zeta_y^\prime$ will have negligible effect on the fine structure levels and state coupling and thus can also be ignored. Defining the effective strain field $\vec{\sigma} = \sigma_x\vec{x}+\sigma_y\vec{y}+\sigma_z\vec{z} = \zeta_x/d_\perp\vec{x}+\zeta_y/d_\perp\vec{y}+\zeta_z/d_\parallel\vec{z}$, the effective matrix representation of $\hat{V}_{str}$ becomes
\begin{eqnarray}
V_\mathrm{str} = \left(\begin{array}{ccc}
0 & 0 & 0 \\
0 & d_\parallel \sigma_z+d_\perp \sigma_x  & -d_\perp \sigma_y \\
0 & -d_\perp \sigma_y & d_\parallel \sigma_z-d_\perp \sigma_x \\
\end{array}\right)
\end{eqnarray}
and it is clear that the strain field can be treated as an additional effective local electric field $\vec{\sigma}$ at the center. Hence, by defining the total effective electric field $\vec{\Pi} = \vec{E}+\vec{\sigma}$, the interaction of the center with both strain and electric fields can be expressed in the spin-Hamiltonian form $\hat{V}_{el}+\hat{V}_{str} = \frac{1}{\hbar^2}d_\parallel \Pi_z S_z^2-\frac{1}{\hbar^2}d_\perp \Pi_x(S_x^2-S_y^2)+\frac{1}{\hbar^2}d_\perp \Pi_y(S_xS_y+S_yS_x)$.

\subsection{The complete spin-Hamiltonian}

Including the descriptions of the spin-spin zero-field splitting and interactions with magnetic, electric and strain fields obtained in the previous subsections, the complete electronic spin-Hamiltonian of the ground state spin becomes
\begin{eqnarray}
\hat{H}_{gs} &=& \frac{1}{\hbar^2}(D_{gs}+d_\parallel \Pi_z) S_z^2+\frac{\mu_B}{\hbar}\vec{S}\cdot\bar{g}\cdot\vec{B}\nonumber \\
&&-\frac{1}{\hbar^2}d_\perp \Pi_x(S_x^2-S_y^2)+\frac{1}{\hbar^2}d_\perp \Pi_y(S_xS_y+S_yS_x) \nonumber \\ \label{eq:spinhamiltonian}
\end{eqnarray}
which in the spin basis $\{S_{A_2},-S_{E,y},S_{E,x}\}$ associated with the ground triplet spin-orbit states, has the matrix representation
\begin{eqnarray}
\hat{H}_{gs} = \left(\begin{array}{ccc}
0 & i{\cal B}_y & -i{\cal B}_x \\
-i{\cal B}_y & {\cal D}+{\cal E}_x & -i{\cal B}_z-{\cal E}_y \\
i{\cal B}_x & i{\cal B}_z-{\cal E}_y  & {\cal D}-{\cal E}_x \\
\end{array}\right)
\end{eqnarray}
where ${\cal D} = D_{gs}+d_\parallel\Pi_z$, ${\cal B}_z = \mu_Bg_\parallel B_z$, ${\cal E}_k = d_\perp \Pi_k$, ${\cal B}_k = \mu_Bg_\perp B_k$, and $k$=$x$,$y$. This final matrix representation provides the simplest description of the dependence of the ground state spin on the six independent electric-magnetic-strain field parameters $({\cal D}, {\cal E}_x,{\cal E}_y,\vec{{\cal B}})$ and will be used to obtain the spin solution in part II of this paper series.\cite{partII}

\section{Nuclear hyperfine structure}

The nuclear hyperfine interaction $\hat{V}_{hf} = \hat{V}_{mhf}+\hat{V}_{ehf}$ between the center's electrons and the nuclei of the crystal lattice has magnetic $\hat{V}_{mhf}$ and electric $\hat{V}_{ehf}$ components. The magnetic hyperfine component accounts for the interactions of the electronic spin and orbital magnetic moment with the nuclear spins of the lattice. Since it was found in the previous section that the ground triplet has no orbital magnetic moment at zero-order in the spin coupling coefficients, the interaction between the orbital magnetic moment and the nuclear spins maybe ignored. The electric hyperfine component approximately accounts for the finite size of the nuclei through the interaction of the center's electrons with the electric quadrupole moments of the finite charge distributions of the nuclei. In an isotopically pure $^{12}\mathrm{C}$ crystal, only the $^{14}\mathrm{N}$ nucleus of the center will have a non-zero nuclear spin ($I=1$) and a non-zero electric quadrupole moment. \cite{stoneham} Consequently, in an isotopically pure crystal the magnetic and electric hyperfine interactions of the NV$^-$ center are described by the potentials \cite{stoneham}
\begin{eqnarray}
\hat{V}_{mhf} & = &  C_{mhf}\sum_i \left(4\pi\delta(\vec{r}_{iN})-\frac{1}{|\vec{r}_{iN}|^3}\right)\vec{s}_i\cdot\vec{I} \nonumber \\
&& +\frac{3(\vec{s}_i\cdot\vec{r}_{iN})(\vec{r}_{iN}\cdot\vec{I})}{|\vec{r}_{iN}|^5}\nonumber \\
\hat{V}_{ehf} & = & \frac{1}{2Z_N}\sum_{\alpha,a,b}\left(
\sum_i\left.\frac{\partial^2\hat{V}_{e}(\vec{r}_i)}{\partial R_{N,a}\partial R_{N,b}}\right|_{\vec{R}_0}\right. \nonumber \\
&& \left.+\left.\frac{\partial^2V_{l}}{\partial R_{N,a}\partial R_{N,b}}\right|_{\vec{R}_0}\right) u_{\alpha,a}u_{\alpha,b}
\end{eqnarray}
where $C_{mhf}=\mu_B\mu_Ng_eg_N\frac{\mu_0}{4\pi\hbar^2}$, $\mu_N$ is the nuclear magneton, $g_N = 0.40356$  is the $^{14}\mathrm{N}$ nuclear g-factor,\cite{MQM} $\vec{I}$ is the spin operator of the $^{14}\mathrm{N}$ nucleus, $\vec{r}_{iN} = \vec{R}_N-\vec{r}_i = x_{iN}\vec{x}+y_{iN}\vec{y}+z_{iN}\vec{z}$, $\vec{R}_N= R_{N,x}\vec{x}+R_{N,y}\vec{y}+R_{N,z}\vec{z}$ is the position of the $^{14}\mathrm{N}$ nucleus, $\hat{V}_{e}$ is the Coulomb interaction potential of the center's electrons with the $^{14}\mathrm{N}$ nucleus, $V_{l}$ is the Coulomb interaction potential of the lattice electrons and $^{12}\mathrm{C}$ nuclei with the $^{14}\mathrm{N}$ nucleus, $Z_N=7$ is the relative charge of the $^{14}\mathrm{N}$ nucleus, and $u_{\alpha,a}$ and $u_{\alpha,b}$ are the components of the displacement of the $\alpha^{th}$ $^{14}\mathrm{N}$ proton from $\vec{R}_N$ in the $a,b = x,y,z$ directions.

The magnetic hyperfine interaction can be written as a sum of terms containing rank two orbital tensor operators $\hat{V}_{mhf} = \sum_i\vec{s}_i\cdot[\bar{A}_{A_1}(\vec{r_{iN}})+\bar{A}_{E,x}(\vec{r_{iN}})+\bar{A}_{E,y}(\vec{r_{iN}})]\cdot\vec{I}$. By applying the matrix representations of table \ref{tab:spinandorbitaloperator}, it is clear that only the term containing $\bar{A}_{A_1}(\vec{r_{iN}})$ contributes at zero-order in the spin coupling coefficients. Given this result, $\hat{V}_{mhf}$ can be written in the spin-Hamiltonian form $\hat{V}_{mhf} = \vec{S}\cdot\bar{A}\cdot\vec{I}$, where
\begin{eqnarray}
\bar{A} = \left(\begin{array}{ccc}
A_\perp & 0 & 0 \\
0 & A_\perp & 0 \\
0 & 0 & A_\parallel \\
\end{array}\right), \nonumber
\end{eqnarray}
$A_\parallel = f_{A_1}+2a_{A_1}$ is the axial magnetic hyperfine parameter, $A_\perp = f_{A_1}-a_{A_1}$ is the non-axial magnetic hyperfine parameter, $f_{A_1} =C_{mhf}4\pi\rdm{e}{\delta(\vec{r}_{iN})}{e} $ is the Fermi contact contribution,  and $a_{A_1} = \frac{1}{2}C_{mhf}\rdm{e}{\frac{1}{|r_{iN}|^3}\left(\frac{3z_{iN}^2}{|\vec{r}_{iN}|^2}-1\right)}{e}$ is the dipolar contribution. Measured values of the magnetic hyperfine parameters are contained in table \ref{tab:ESRparameters} and, although they differ in sign and magnitude, conclusively show that both the Fermi contact and dipolar contributions must be non-zero. For example, using the values obtained in Ref. \onlinecite{gfactor}, $f_{A_1}/h = -2.51$ MHz and $a_{A_1}/h = 187$ kHz.

The expression for the Fermi contact contribution can be simplified further to $f_{A_1} = C_{mhf}4\pi|e_x(\vec{R}_N)|^2= C_{mhf}4\pi|e_y(\vec{R}_N)|^2$. As the $e$ MOs transform as the $E$ irreducible representation of the $C_{3v}$ group, they are by definition zero at any point along the axial symmetry axis of the center, and since the equilibrium position of the $^{14}\mathrm{N}$ nucleus is on the axial symmetry axis, the Fermi contact contribution vanishes if the $^{14}\mathrm{N}$ is fixed at its equilibrium position. To account for the non-zero Fermi contact contribution in the molecular model, the vibrational wavefunction $\chi_N(\vec{R}_N)$ of the $^{14}\mathrm{N}$ must be considered, in which case the expression for the Fermi contact contribution becomes
\begin{eqnarray}
f_\mathrm{vib} = C_{mhf}4\pi\int|e_x(\vec{R}_N)|^2|\chi_N(\vec{R}_N)|^2d^3R_N
\end{eqnarray}
A similar vibrationally corrected expression for the dipolar contribution can also be defined.

The magnetic polarization of the $^{14}\mathrm{N}$ core electrons associated with the $m_s = \pm1$ states of the center will provide a negative Fermi contact contribution $f_{core}$ \cite{galihyperfine,gfactor} in addition to the positive contribution arising from the vibrational motion of the nucleus. However, the description of $f_{core}$ is beyond the molecular model in its current formulation, as only the interactions of the bound valence electrons are considered in the model. An \textit{ab initio} study \cite{galihyperfine} has yielded $f_{core} \approx -1.7$ MHz, but no \textit{ab initio} calculation of the vibrationally corrected $f_\mathrm{vib}$ has been conducted to date. Although $f_{vib}$ is likely to be much smaller than $f_{core}$, the calculation of $f_\mathrm{vib}$ will provide a more complete \textit{ab initio} model of the magnetic hyperfine parameters. Likewise, an extension of the molecular model to describe $f_{core}$ will also provide further insight into the interactions between the nucleus and the bound electrons.

The electronic component of the electric hyperfine interaction $\hat{V}_{ehf}$ can be written as a sum of products of orbital electronic and nuclear tensor operators. Analogous to the interaction of the ground state spin with electric and strain fields, at zero-order in the spin coupling coefficients only the terms of $\hat{V}_{ehf}$ that contain $A_1$ orbital electronic tensor operators will have non-zero matrix elements, and these matrix elements will be diagonal and identical for each spin state. The symmetry of the lattice ensures that only similar terms from the lattice electron and internuclear contribution are also non-zero. Consequently, the electric hyperfine interaction within the ground state spin reduces to
\begin{eqnarray}
\hat{V}_{ehf} = \frac{q_{z}}{4Z_N}\sum_\alpha 3u_{\alpha,z}^2-|\vec{u}_\alpha|^2
\end{eqnarray}
where $q_{z} = 2\rdm{e}{\partial^2\hat{V}_{e}/\partial R_{N,z}^2|_{\vec{R}_0}}{e}+2\rdm{a_1}{\partial^2\hat{V}_{e}/\partial R_{N,z}^2|_{\vec{R}_0}}{a_1}+\partial^2V_{l}/\partial R_{N,z}^2|_{\vec{R}_0}$ is proportional to the axial gradient of the net axial electric field at the $^{14}\mathrm{N}$ nucleus generated by all of the electrons and other nuclei.

As per standard practice, the nuclear quadrupole operator $\sum_\alpha 3u_{\alpha,z}-|\vec{u}_\alpha|^2$ can be replaced by a nuclear spin operator $Q_{z}I_z^2/\hbar^2$ of the same symmetry through the definition of the proportionality constant $Q_z =\bra{1,1}\sum_\alpha 3u_{\alpha,z}-|\vec{u}_\alpha|^2\ket{1,1}$ using the $I_z$ eigenstate $\ket{I=1,m_I =1}$.\cite{quadrupole} $Q_z$ therefore quantifies the difference in the axial anisotropy of the nuclear quadrupole moment between the $m_I=\pm1$ and $m_I=0$ states. The final form of the effective electric hyperfine interaction is then $\hat{V}_{ehf} = PI_z^2/\hbar^2$, where $P=q_{z}Q_{z}/4Z_N$ is the nuclear quadrupole parameter contained in table \ref{tab:ESRparameters}.

Combining the expressions obtained for the magnetic and electric hyperfine interactions, the zero-field spin-Hamiltonian of the ground state spin including nuclear hyperfine interaction is
\begin{eqnarray}
\hat{H}_{hf} &=& \frac{1}{\hbar^2}\left[D_{gs}S_z^2+A_\parallel S_zI_z\right.\nonumber \\
&& \left.+A_\perp(S_xI_x+S_yI_y)+PI_z^2\right]
\end{eqnarray}
The approximate solutions of $\hat{H}_{hf}$ can be obtained by constructing the hyperfine states of the ground triplet in a similar manner to the earlier construction of the electronic spin-orbit states by defining linear combinations of products of electronic spin-orbit and nuclear spin states that have definite $C_{3v}$ symmetry. Given that the symmetrised nuclear spin states of the $^{14}\mathrm{N}$ nucleus in terms of the $I_z$ eigenstates $\{\ket{I,m_I}\}$ are $I_{A_2} = \ket{1,0}$, $I_{E,x} = \frac{-i}{\sqrt{2}}(\ket{1,1}+\ket{1,-1})$ and $I_{E,y} = \frac{-1}{\sqrt{2}}(\ket{1,1}-\ket{1,-1})$, the symmetrised hyperfine states $\Psi_{n,j,k}$ are
\begin{eqnarray}
&&\Psi_{1,E,x} = \Phi_{1,A_1}^{so}I_{E,x}, \
\Psi_{1,E,y} = \Phi_{1,A_1}^{so}I_{E,y} \nonumber \\
&& \Psi_{2,A_2} = \Phi_{2,A_1}^{so}I_{A_2} \nonumber \\
&&\Psi_{3,E,x} = \frac{1}{\sqrt{2}}(\Phi_{2,E,x}^{so}I_{E,x}-\Phi_{2,E,y}^{so}I_{E,y}) \nonumber
\end{eqnarray}
\begin{eqnarray}
&&\Psi_{3,E,y} = \frac{-1}{\sqrt{2}}(\Phi_{2,E,x}^{so}I_{E,y}-\Phi_{2,E,y}^{so}I_{E,x}) \nonumber\\
&&\Psi_{4,A_1} = \frac{1}{\sqrt{2}}(\Phi_{2,E,x}^{so}I_{E,x}+\Phi_{2,E,y}^{so}I_{E,y}) \nonumber \\
&&\Psi_{5,A_2} = \frac{1}{\sqrt{2}}(\Phi_{2,E,x}^{so}I_{E,y}-\Phi_{2,E,y}^{so}I_{E,x}) \nonumber \\
&&\Psi_{6,E,x} = -\Phi_{2,E,y}^{so}I_{A_2}, \ \Psi_{6,E,y} = \Phi_{2,E,x}^{so}I_{A_2}
\end{eqnarray}

\begin{figure*}[hbtp]
\begin{center}
\mbox{
\subfigure[]{\includegraphics[width=0.85\columnwidth] {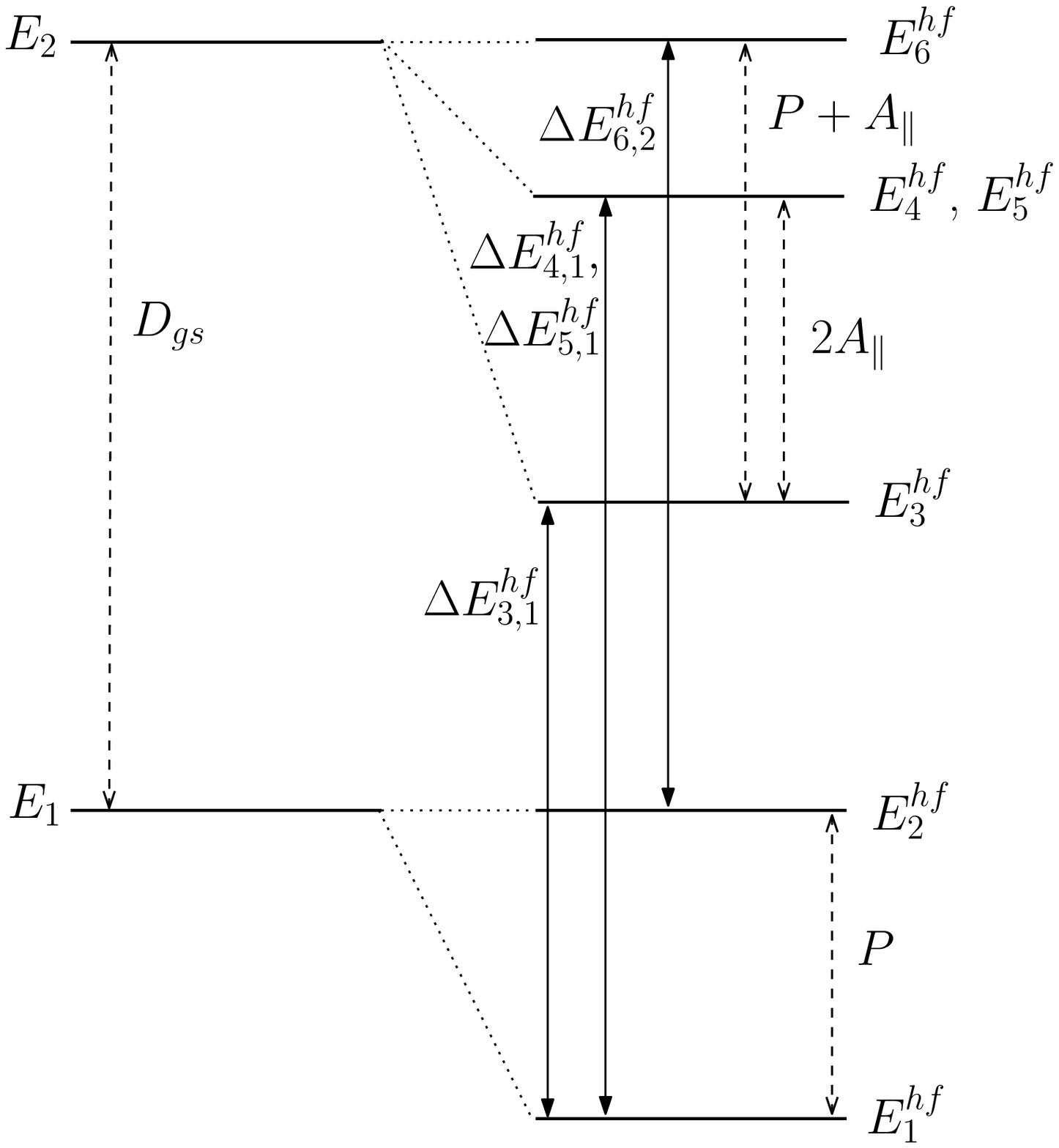}}
\subfigure[]{\includegraphics[width=1.0\columnwidth] {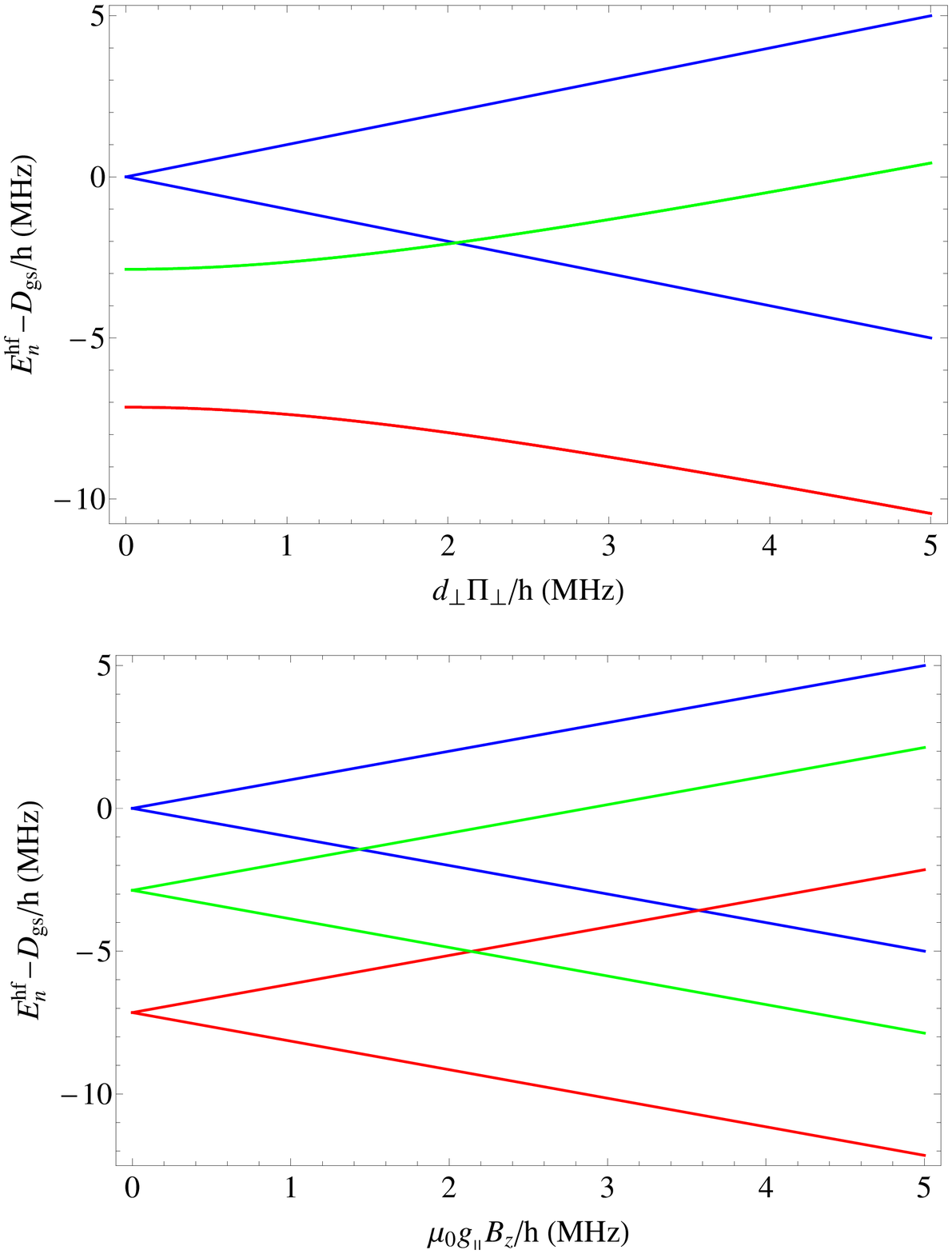}}
}
\caption{(color online) (a) The hyperfine structure of the ground triplet. The level splittings are indicated by dashed arrows and the observed values for the hyperfine parameters $A_\parallel$ and $P$ are contained in table \ref{tab:ESRparameters}. The ordering of the hyperfine levels is based upon the parameters measured in Ref. \onlinecite{gfactor}. The allowed magnetic transitions between levels of the same nuclear spin projection are indicated by solid arrows and labeled by their transition energy $\Delta E_{n,m}^{hf}$. (b) The splittings of the $m_s = \pm1$ hyperfine levels in the presence of  non-axial electric-strain fields of magnitude $\Pi_\perp$ (upper) and  axial magnetic fields $B_z$ (lower). $m_I=0$ hyperfine levels are colored blue and $m_I = \pm1$ hyperfine levels are colored red and green. The hyperfine parameters of Ref. \onlinecite{gfactor} have been used to calculate the splittings.}
\label{fig:hyperfinestructure}\label{fig:hyperfinesplittings}
\end{center}
\end{figure*}

The matrix representation of the zero-field Hamiltonian in the basis of hyperfine states $\{\Psi_{2,A_2},$ $\Psi_{6,E,x},$ $\Psi_{6,E,y},$ $\Psi_{1,E,x},$ $\Psi_{1,E,y},$ $\Psi_{3,E,x},$ $\Psi_{3,E,y},$ $\Psi_{4,A_1},$ $\Psi_{5,A_2} \}$ is
\begin{widetext}
\begin{eqnarray}
H_{hf} =
\left(
\begin{array}{ccccccccc}
 0 & 0 & 0 & 0 & 0 & 0 & 0 & 0 & \sqrt{2}A_\perp \\
 0 & D_{gs} & 0 & -A_\perp & 0 & 0 & 0 & 0 & 0 \\
 0 & 0 & D_{gs} & 0 & -A_\perp & 0 & 0 & 0 & 0 \\
 0 & -A_\perp & 0 & P & 0 & 0 & 0 & 0 & 0 \\
 0 & 0 & -A_\perp & 0 & P & 0 & 0 & 0 & 0 \\
 0 & 0 & 0 & 0 & 0 & D_{gs}+h_+ & 0 & 0 & 0 \\
 0 & 0 & 0 & 0 & 0 & 0 & D_{gs}+h_+ & 0 & 0 \\
 0 & 0 & 0 & 0 & 0 & 0 & 0 & D_{gs}+h_- & 0 \\
 \sqrt{2}A_\perp & 0 & 0 & 0 & 0 & 0 & 0 & 0 & D_{gs}+h_- \\
\end{array}
\right) \nonumber \\
\end{eqnarray}
\end{widetext}
where $h_\pm = P\pm A_\parallel$. Note that the above basis of hyperfine states has been grouped into states of the same nuclear spin projection. The energies $E_n^{hf}$ correct to first-order in nuclear hyperfine interactions can be easily inferred: $E_1^{hf} = P$, $E_2^{hf} = 0$, $E_3^{hf} = D_{gs}+P+A_\parallel$, $E_4^{hf}  = E_5^{hf} = D_{gs}+P-A_\parallel$,  $E_6^{hf} = D_{gs}$; and the corresponding hyperfine structure is depicted in Fig. \ref{fig:hyperfinestructure}. Since $A_\perp$ is observed to satisfy $A_\perp \ll D_{gs}$, \cite{xing,gfactor} the non-axial magnetic hyperfine parameter that couples hyperfine states of different nuclear spin projection will have a negligible effect on the zero-field hyperfine structure and state couplings.

The matrix representation of the electronic interaction with electric, magnetic and strain fields in the hyperfine basis is
\begin{widetext}
\begin{eqnarray}
V_\mathrm{mag}+V_\mathrm{el}+V_\mathrm{str} =
\left(
\begin{array}{ccccccccc}
 0 & i{\cal B}_x & i{\cal B}_y & 0 & 0 & 0 & 0 & 0 & 0 \\
 -i{\cal B}_x & d_\parallel \Pi_z-{\cal E}_x & {\cal E}_y-i{\cal B}_z & 0 & 0 & 0 & 0 & 0 & 0 \\
 -i{\cal B}_y & i{\cal B}_z+{\cal E}_y & d_\parallel \Pi_z+{\cal E}_x & 0 & 0 & 0 & 0 & 0 & 0 \\
 0 & 0 & 0 & 0 & 0 & \frac{i{\cal B}_y}{\sqrt{2}} & \frac{i{\cal B}_x}{\sqrt{2}} & \frac{i{\cal B}_y}{\sqrt{2}} & \frac{i{\cal B}_x}{\sqrt{2}} \\
 0 & 0 & 0 & 0 & 0 & \frac{i{\cal B}_x}{\sqrt{2}} & -\frac{i{\cal B}_y}{\sqrt{2}} & -\frac{i{\cal B}_x}{\sqrt{2}} & \frac{i{\cal B}_y}{\sqrt{2}} \\
 0 & 0 & 0 & -\frac{i{\cal B}_y}{\sqrt{2}} & -\frac{i{\cal B}_x}{\sqrt{2}} & d_\parallel \Pi_z & i{\cal B}_z & {\cal E}_x & {\cal E}_y \\
 0 & 0 & 0 & -\frac{i{\cal B}_x}{\sqrt{2}} & \frac{i{\cal B}_y}{\sqrt{2}} & -i{\cal B}_z & d_\parallel \Pi_z & {\cal E}_y & -{\cal E}_x \\
 0 & 0 & 0 & -\frac{i{\cal B}_y}{\sqrt{2}} & \frac{i{\cal B}_x}{\sqrt{2}} & {\cal E}_x & {\cal E}_y & d_\parallel \Pi_z & i{\cal B}_z \\
 0 & 0 & 0 & -\frac{i{\cal B}_x}{\sqrt{2}} & -\frac{i{\cal B}_y}{\sqrt{2}} & {\cal E}_y & -{\cal E}_x & -i{\cal B}_z & d_\parallel \Pi_z
\end{array}
\right) \nonumber \\
\end{eqnarray}
\end{widetext}
The matrix representation demonstrates that if the much smaller interaction of the nuclear spin with the fields is ignored, the fields do not couple states of different nuclear spin projection. Furthermore, by comparing the above matrix representation with that of $\hat{H}_{gs}$, the representation also demonstrates that the hyperfine states of nuclear spin projection $m_I = 0$ (upper $3\times3$ diagonal block) interact with the fields in a manner similar to the electronic spin-orbit states discussed in the previous section, whereas the $m_I = \pm1$ hyperfine states (lower $6\times6$ diagonal block) interact differently in the weak field limit, where the fields induce shifts comparable to the hyperfine splittings. For example, as depicted in Fig. \ref{fig:hyperfinesplittings}, the ($m_s = \pm1$, $m_I = 0$) states split linearly in the presence of a non-axial electric-strain field, whereas the non-degenerate ($m_s = \pm1$, $m_I = \pm1$) states repel quadratically in the presence of non-axial electric-strain fields that satisfy ${\cal E}_\perp < 2A_\parallel$. Note that in the large field limit, where the fields induce shifts much larger than the hyperfine splittings, both sets of hyperfine states behave approximately analogous to the spin-orbit states.

The allowed magnetic transitions are depicted in Fig. \ref{fig:hyperfinestructure} and indicate that in the absence of static fields, there will exist three lines in the hyperfine spectra with energies $\Delta E_{3,1}^{hf} = D_{gs}+A_\parallel$, $\Delta E_{4,1}^{hf} = \Delta E_{5,1}^{hf} = D_{gs}-A_\parallel$ and $\Delta E_{6,2}^{hf} = D_{gs}$ in agreement with observation. \cite{loubser} The central hyperfine line therefore corresponds to transitions between $m_I = 0$ states and the lower and higher energy lines correspond to transitions between $m_I = \pm1$ states. Consequently, due to the different interactions of the $m_I = 0$ and $m_I = \pm1$ states in the weak static field limit, the central hyperfine line will depend differently on the static fields compared to the lower and higher energy lines. These differences in the dependence of the hyperfine lines in the weak field limit was used in the recent electric field sensing demonstration, \cite{efield} where the magnetic field was precisely aligned in the non-axial direction ($B_z = 0$) in the presence of a non-axial electric-strain field by observing the splitting of the $m_I = \pm1$ hyperfine lines whilst the measurement of the electric field was conducted by observing the linear splitting of the central $m_I = 0$ hyperfine line. Hence, it is clear that the hyperfine structure of the ground state spin and its more complicated interactions with electric, magnetic and strain fields, is an important consideration for applications of the spin that operate in the weak field limit.

\section{Conclusion}

In this article, the theory of the ground state spin has for the first time been fully developed using the molecular model of the center in order to provide detailed explanations for the spin's fine and hyperfine structures and its interactions with electric, magnetic and strain fields. Given these explanations, an effective spin-Hamiltonian that describes the electronic states in the high field limit and the $m_I = 0$ subset of hyperfine states in the low field limit was derived. The explanations also allowed the correlation of the properties of the ground state spin with the other properties of the center and provided explicit expressions for the key parameters of spin in terms of the center's MOs. Hence, this work has identified the critical parameters that need to be pursued by future experimental and \textit{ab initio} studies. Furthermore, this work has also provided the essential theoretical understanding of this remarkable spin that can be used to model the spin in its ground-breaking quantum metrology and QIP applications.

\begin{acknowledgments}
This work was supported by the Australian Research Council under the
Discovery Project scheme (DP0986635 and DP0772931), the EU commission (ERC grant SQUTEC), Specific Targeted Research Project  DIAMANT and the integrated project SOLID. F.D. wishes to acknowledge the Badenwurttenberg Stiftung Internat.
Spittenforschung II MRI.
\end{acknowledgments}

\end{document}